\documentclass[twocolumn,aps,showpacs]{revtex4}

\usepackage{amsmath}
\usepackage{amssymb}
\usepackage{graphicx}
\usepackage{dcolumn}

\begin{document}

\title{Theory of acoustic-phonon assisted magnetotransport \\
in 2D electron systems at large filling factors}

\author{O. E. Raichev}
\affiliation{Institute of Semiconductor Physics, National Academy of Sciences of Ukraine, 
Prospekt Nauki 41, 03028, Kiev, Ukraine}
\date{\today}

\begin{abstract}
A microscopic theory of the phonon-induced resistance oscillations in weak 
perpendicular magnetic fields is presented. The calculations are based on the 
consideration of interaction of two-dimensional electrons with three-dimensional 
(bulk) acoustic phonons and take into account anisotropy of the phonon spectrum 
in cubic crystals. The magnetoresistance is calculated for [001]-grown GaAs 
quantum wells. The results are in agreement with available experimental data. 
Apart from the numerical results, analytical expressions for the oscillating part 
of magnetoresistance are obtained. These expressions are valid in the region of 
high-order magnetophonon resonances and describe the oscillating magnetoresistance 
determined by several groups of phonons polarized along certain high-symmetry 
directions. 

\end{abstract}


\maketitle

\section{Introduction}

In recent years, experimental studies of transport properties of high-mobility 
two-dimensional (2D) electron gas in weak perpendicular magnetic fields have uncovered 
a variety of remarkable quantum phenomena caused by transitions of electrons 
between different Landau levels. Such transitions can lead to oscillations of dissipative 
resistance as a function of the magnetic field. For example, under steady-state 
microwave illumination of the 2D gas the resistance oscillates with a period 
determined by the ratio of the radiation frequency to the cyclotron frequency 
$\omega_{c}$. This phenomenon is known as the microwave-induced resistance 
oscillations (MIRO).$^1$ For a sufficiently high radiation power the MIRO minima evolve into 
the intervals of magnetic field where the dissipative resistance vanishes.$^{2,3}$ 
Next, it was found that an increase in the electric current passing though the 2D layer 
substantially reduces the resistance$^{4,5}$ and leads to oscillations of the resistance 
as a function of either the magnetic field or the current.$^{4,6,7}$ Such oscillations are 
controlled by the ratio of a characteristic energy, which is defined as the drop of 
the Hall electric field across the classical cyclotron diameter, to the cyclotron 
energy. This phenomenon has been called the Hall field-induced resistance oscillations 
(HIRO). It is also described in terms of Zener tunneling between Landau levels. 
Finally, in the systems with two occupied 2D subbands the resistivity oscillates 
as a function of the ratio of subband separation energy to the cyclotron 
energy owing to coupling of the subbands via intersubband scattering. This
phenomenon, called the magneto-intersubband oscillations (MISO), is important 
in the systems with small subband separation, such as the double quantum 
wells.$^{8,9}$ All these kinds of oscillations are insensitive to positions of the 
Landau levels with respect to the Fermi energy. Therefore, unlike the Shubnikov-de 
Haas oscillations, they are not damped exponentially with increasing temperature. 

The oscillatory phenomena described above require the presence of microwave illumination, 
strong Hall field, or intersubband coupling to enable resonant transitions between different Landau 
levels via elastic scattering of electrons by impurities or other static imperfections of 
the system. The elastic scattering provides the main contribution to resistivity at low 
temperatures. In high-mobility samples, however, inelastic scattering of electrons 
by acoustic phonons also contributes to transport, and can exceed the impurity-scattering 
contribution at the temperatures of several K. Therefore, acoustic-phonon-assisted transitions 
of electrons between Landau levels become important. Interestingly enough, these transitions 
in 2D case possess a resonant property and, for this reason, also lead to oscillations of 
the resistance. Oscillations of this origin have been discovered recently.$^{10}$ An analysis 
of experimental data has suggested that the resistivity acquires an oscillating contribution 
whose periodicity is determined by the resonant condition 
\begin{equation}
2 p_F s= n \hbar \omega_c, 
\end{equation}
where $p_F$ is the Fermi momentum of electrons, $s$ is the constant of the order of sound 
velocity, and $n$ is an integer.$^{10}$ This phenomenon has been called the phonon-induced 
resistance oscillations (PIRO). Similar oscillations have been observed later in 
the phonon drag thermal power measurements.$^{11}$ 

Since the quantity in the left-hand side of Eq. (1) is a characteristic phonon energy, 
the PIRO can be viewed as magnetophonon oscillations. For the case of electron 
scattering by optical phonons, the magnetophonon oscillations are known for a long 
time$^{12}$ and have been observed both in bulk and 2D$^{13}$ systems. The special 
feature of PIRO, in contrast to the oscillations owing to scattering by optical 
phonons, is the dependence of magnetophonon resonance on the electron momentum. By 
emitting or absorbing acoustic phonons, the electrons jump between the Landau levels, 
and the maximum probability of such transitions is realized under electron backscattering 
condition, when the phonon wave number $q$ reaches its maximum value $2p_F/\hbar$ in the 
2D plane. Initially,$^{10}$ this resonant property of the electron-phonon scattering 
was explained by involving the interface phonon model. Later, by analyzing phonon-assisted 
transitions between Landau levels, it was found$^{11}$ that the bulk phonon model 
gives similar results. It is worth noting that the maximum probability of the 
backscattering processes ($q=2p_F/\hbar$) is not related to the presence of the magnetic 
field. It reflects a fundamental property of the kinematics of two-dimensional electron 
scattering by three-dimensional acoustic modes with frequency $\omega_{\bf q}=s 
\sqrt{q_{\bot}^2+q^2_z}$, where $q_{\bot}$ and $q_z$ are the in-plane and out-of-plane 
components of the phonon wave vector. It can be shown that the scattering probability, 
as a function of the phonon frequency, has a logarithmic singularity at the point of 
transition from the region $\omega_{\bf q} < 2p_Fs/\hbar$, when pure 2D scattering ($q_z=0$) 
is possible, to the region $\omega_{\bf q} > 2p_Fs/\hbar$, when emission of three-dimensional 
phonons (with finite $q_z$) becomes necessary. 

The systematic studies of the PIRO are now under way.$^{14-17}$ However, current 
understanding of this phenomenon is far from complete. The characteristic phonon 
velocity $s$ entering Eq. (1) varies in different experiments$^{10,14,15,17}$ in 
the range from 2.9 to 5.9 km/s, and the origin of such variations is not clear.  
The problem of the phase of the oscillations has not been discussed. The dependence 
of the amplitude of the oscillations on temperature and magnetic field has not been 
investigated in detail, though general features of the experimental dependence have 
been successfully related$^{17}$ to the behavior of the density of electron states in 
magnetic field. The main difficulty for interpretation of the experimental magnetoresistance 
is the complicated nature of acoustic-phonon modes. Even in the case when the influence of 
interfaces on the phonon spectrum is neglected (bulk phonon approximation), one should 
take into account three different phonon branches characterized by anisotropic 
(i.e., dependent on the direction of phonon momentum) velocities.
 
In this paper, the theory of PIRO is developed within the bulk phonon approximation. 
Both deformation-potential and piezoelectric mechanisms of electron-phonon interaction 
are included into consideration. The anisotropy of phonon spectrum is taken into 
account. The calculations are carried out for GaAs quantum well layers grown in the 
[001] crystallographic direction. Apart from the numerical results, approximate analytical 
expressions for the oscillating part of the phonon-induced resistivity are presented. 
The results are compared with available experimental data for two kinds of GaAs
layers with distinctly different densities and mobilities of 2D electrons.

The paper is organized as follows. Section II contains general description of the 
resistivity of 2D electrons interacting with acoustic phonons in a weak magnetic 
field. Section III includes both analytical and numerical results for the resistivity, 
their comparison with experiment, and discussion. The summary and conclusions are 
given in the last section.   
 
\section{General formalism}

Experimental investigations of phonon-induced resistance oscillations are 
carried out at low temperatutes $T$ satisfying the condition of degenerate 
electron gas, $T \ll \varepsilon_F$, where $\varepsilon_F$ is the Fermi energy. 
Magnetic fields are weak enough to have many Landau levels populated, 
$\hbar \omega_c \ll \varepsilon_F$, so the electron transport possesses some 
quasiclassical features (for example, the electron momentum is still a good 
quantum number). In the same time, the high quality of the samples ensures that the 
Landau quantization is not suppressed by scattering at such low magnetic fields. 
This means that the ratio of the cyclotron frequency $\omega_c$ to inverse quantum 
lifetime $1/\tau_q$ is not too small. Next, since in the high-mobility samples the 
ratio of transport time to quantum lifetime is large, the oscillations are 
observed in the regime of classically strong magnetic fields, when the cyclotron 
frequency is much larger than the inverse transport time $1/\tau_{tr}$. The 
in-plane transport in this case can be viewed as hopping between cyclotron orbit 
centers, and the dissipative conductivity $\sigma_d$ is represented in the form
\begin{equation}
\sigma_{d}= \frac{2 e^2}{T L^2} \sum_{\delta \delta^{\prime}}
f(\varepsilon_{\delta}) [1- f(\varepsilon_{\delta^{\prime}}) ] \nu_{\delta
\delta^{\prime}} (X_{\delta} - X_{\delta^{\prime}})^2/2,
\end{equation}
where $\delta$ is the index of a quantum state of electron, $X_{\delta}$ is 
the coordinate of the cyclotron orbit center in the direction of motion, 
$f(\varepsilon)$ is the equilibrium Fermi-Dirac distribution function, 
$L^2$ is the normalization area, and $\nu$ is the total scattering rate, which 
is represented as a sum of  impurity and phonon scattering contributions: 
$\nu_{\delta \delta^{\prime}}= \nu^{im}_{\delta \delta^{\prime}} +\nu^{ph}_{\delta 
\delta^{\prime}}$. Assuming interaction of electrons with equilibrium  
phonons, the latter contribution is written as
\begin{eqnarray}
\nu^{ph}_{\delta \delta^{\prime}}= \frac{2 \pi}{\hbar} \sum_{\lambda} \int \frac{d {\bf q}}{(2 \pi)^3}
C_{\lambda {\bf q}} | \langle \delta| e^{i {\bf q \cdot r}}|
\delta^{\prime}\rangle |^2 \left[ (N_{\lambda {\bf q}}+1) \right. \nonumber \\
\left. \times \delta(\varepsilon_{\delta} - \varepsilon_{\delta^{\prime}}
- \hbar \omega_{\lambda {\bf q}}) +  N_{\lambda {\bf q}} \delta(\varepsilon_{\delta} -
\varepsilon_{\delta^{\prime}} + \hbar \omega_{\lambda {\bf q}})   \right],
\end{eqnarray}
where $\lambda$, ${\bf q}=(q_x,q_y,q_z)$, and $\omega_{\lambda {\bf q}}$ are 
the phonon mode index, wave vector, and frequency, respectively. Next, 
$N_{\lambda {\bf q}}=[\exp(\hbar \omega_{\lambda {\bf q}}/T)-1]^{-1}$ is the 
Planck distribution function. The interaction is described by the function
$C_{\lambda {\bf q}}$, which includes both deformation-potential and piezoelectric 
coupling of electrons to crystal vibrations. Considering only acoustic modes in 
cubic crystals, the general form of this function is written as follows:
\begin{eqnarray}
C_{\lambda {\bf q}}=\frac{\hbar}{2 \rho \omega_{\lambda {\bf q}}} 
\biggl[ {\cal D}^2 \sum_{ij} {\rm e}_{\lambda {\bf q} i} {\rm e}_{\lambda {\bf q} j} q_i q_j  \nonumber \\
+\frac{(eh_{14})^2}{q^4} \sum_{ijk,i'j'k'}\kappa_{ijk} \kappa_{i'j'k'} {\rm e}_{\lambda {\bf q} k} 
{\rm e}_{\lambda {\bf q} k'} q_i q_j q_{i'} q_{j'} \biggr], 
\end{eqnarray}
where ${\cal D}$ is the deformation potential constant, $h_{14}$ is the piezoelectric 
coupling constant, and $\rho$ is the material density. The sums are taken over 
Cartesian coordinate indices, ${\rm e}_{\lambda {\bf q} i}$ are the components 
of the unit vector of the mode polarization, and the coefficient $\kappa_{ijk}$ 
is equal to unity if all the indices $i,j,k$ are different and equal to zero 
otherwise. The polarization vectors and the corresponding phonon mode frequencies are 
found from the eigenstate problem
\begin{equation}
\sum_{j} \left[K_{ij}({\bf q})  - \delta_{ij} \rho \omega^2\right] {\rm e}_{\lambda {\bf q} j}=0.
\end{equation}
The dynamical matrix for cubic crystals,
\begin{eqnarray}
K_{ij}({\bf q}) =[(c_{11}-c_{44}) q^2_i + c_{44} q^2] \delta_{ij} \nonumber \\
+ (c_{12}+c_{44}) q_i q_j(1-\delta_{ij}), 
\end{eqnarray}
is written in the elastic approximation and expressed through three elastic 
constants for which the conventional notations are used. Equation (5) has 
three solutions describing a high-energy mode (LA), which becomes purely longitudinal 
for high-symmetry directions ([100],[110], and [111]), and two low-energy 
modes (TA) which become purely transverse for these directions. Very often, 
theoretical studies of transport properties of electrons in solids with cubic 
lattice symmetry are based on the continuum approximation for elastic vibrations, 
when $c_{44}=(c_{11}-c_{12})/2$ and the phonon spectrum is given by three isotropic 
branches including one longitudinal mode and two degenerate transverse modes. Whereas 
this approach gives a reasonably good description of phonon-limited mobilities 
and applies for calculation of many kinetic coefficients, is not sufficient 
for description of phonon-induced resistance oscillations in magnetic 
field, because the anisotropy of the acoustic phonon branches is not weak 
in most semiconductors and becomes essential in evaluation of resonant 
scattering between Landau levels. 

The quantum states $\delta$ in Eq. (2) can be treated as exact eigenstates 
of electrons interacting with randomly distributed impurities in the presence of a
magnetic field. It is convenient to rewrite Eq. (2) through the Green's functions 
of electrons. Using the basis of Landau eigenstates and specifying the growth axis 
of the quantum-well layer as $z$ (i.e., the $[001]$ crystallographic direction), 
one obtains the phonon-induced contribution to the conductivity:
\begin{eqnarray}
\sigma^{ph}_{d}= \frac{e^2 l^2_B}{2 \pi \hbar T} \int_0^{\infty} d q_{\bot} q_{\bot}^3 
\int_{0}^{\infty} \frac{d q_z}{\pi} |\left< 0|e^{iq_z z} |0 \right>|^2 \nonumber \\
\times \int_0^{2 \pi} \frac{d \varphi}{2 \pi}  \sum_{\lambda} C_{\lambda {\bf q}} (N_{\lambda {\bf q}}+1) \sum_{nn'} 
\Phi_{n n^{\prime}}(q_{\bot}^2l^2_B/2)  \nonumber \\
\times \int d \varepsilon f(\varepsilon) [1-f(\varepsilon- \hbar \omega_{\lambda {\bf q}})] 
A_{\varepsilon}(n') A_{\varepsilon - \hbar \omega_{\lambda {\bf q}}}(n), 
\end{eqnarray}  
where $q_{\bot}$ and $\varphi$ are the absolute value and polar angle of the in-plane component 
of phonon wave vector, $n$ are the Landau level numbers, and $l_B$ is the magnetic length. 
The function $\Phi_{n n^{\prime}}(x)= (n!/n^{\prime}!) x^{n^{\prime} -n} e^{-x} 
[L_{n}^{n^{\prime}-n}(x)]^2$, where $L_{n}^{m}(x)$ are the Laguerre polynomials, describes 
scattering in the magnetic field. The squared matrix element of a plane-wave factor, 
$|\left< 0|e^{iq_z z} |0 \right>|^2$, is determined by the confinement potential which 
defines the ground state of 2D electrons, $|0 \rangle$. Applying the model of a deep 
rectangular quantum well of width $d_w$, one can rewrite this squared matrix element as 
$I(q_z d_w/2)$, where $I(x)=(\sin x/x)^2 [1-(x/\pi)^2]^{-2}$. Finally, the spectral 
function $A_{\varepsilon}(n)$, which is equal to the imaginary part of the 
single-electron (advanced) Green's function divided by $\pi$,
characterizes disorder-induced broadening of electron states. In the Born approximation, 
this broadening is conveniently described in terms of the quantum lifetime of electron, 
$\tau_q$. In the absence of impurities (the collisionless limit) $A_{\varepsilon}(n)$ 
is reduced to the delta-function $\delta(\varepsilon-\varepsilon_n)$, where 
$\varepsilon_n=\hbar \omega_c(n+1/2)$ is the Landau quantization energy. 

The dissipative resistivity measured in experiments using Hall bars is related to the conductivity 
$\sigma_d$ according to $\rho_d \simeq \sigma_d/\sigma^2_H$, where $\sigma_H=e^2 n_s/ m \omega_c$ 
is the classical Hall conductivity and $n_s$ is the sheet carrier density. The resistivity is 
represented as a sum of contributions from electron-impurity and electron-phonon scattering: 
\begin{equation}
\rho_d =\rho_{im}+ \rho_{ph},~~~\rho_{im}=\frac{m \nu_{im}}{e^2 n_s},~ \rho_{ph}=\frac{m \nu_{ph}}{e^2 n_s}, 
\end{equation}
where $\nu_{im}$ and $\nu_{ph}$ are the partial scattering rates which depend on the magnetic 
field. Both $\rho_{ph}$ and $\nu_{ph}$ are straightforwardly obtained from $\sigma^{ph}_{d}$ 
of Eq. (7). If the magnetic field is zero, $\nu_{im}$ and $\nu_{ph}$ are reduced to the 
inverse transport times due to electron-impurity and electron-phonon interactions, 
respectively, and Eq. (8) describes the classical (Drude) resistivity. 

Let us consider the case of weak magnetic fields, when many Landau levels are occupied. 
Then one can use smallness of the cyclotron energy $\hbar \omega_c$, phonon 
energy $\hbar \omega_{\lambda {\bf q}}$, disorder-broadening energy $\hbar/\tau_q$, and temperature $T$ 
with respect to the Fermi energy for evaluation of the expression (7). Owing to the sharp 
form of the spectral functions, the contribution into the $n$-sums in Eq. (7) comes from 
a limited number of Landau levels in the vicinity of the Fermi energy, and these sums can 
be approximately calculated with the following result:
\begin{eqnarray}
\nu_{ph}= \frac{2 m}{\hbar^3 T} \int_0^{2 \pi} \frac{d \varphi}{2 \pi} \int_0^{2 \pi} 
\frac{d \theta}{2 \pi} (1-\cos \theta) \nonumber \\
\times \int_{0}^{\infty} \frac{d q_z}{\pi} I \left(\frac{q_z d_w}{2} \right) \sum_{\lambda} 
C_{\lambda {\bf q}} (N_{\lambda {\bf q}}+1) \nonumber \\
\times \int d \varepsilon  f(\varepsilon) [1-f(\varepsilon- \hbar \omega_{\lambda {\bf q}})] 
D({\varepsilon}) D(\varepsilon - \hbar \omega_{\lambda {\bf q}}),
\end{eqnarray}
where $D({\varepsilon})$ is the dimensionless (i.e., normalized to its zero-magnetic-field value) 
density of electron states. The integration over the in-plane phonon wave number $q_{\bot}$ 
is replaced in Eq. (9) with the integration over the scattering angle $\theta$ according 
to $q_{\bot}=2 (p_F/\hbar) \sin(\theta/2)$. This emphasizes that the phonon-assisted transport at large 
filling factors is described within the picture of quasielastic scattering of 2D electrons 
in the vicinity of the Fermi surface. Under the same assumptions, the contribution to the 
resistivity owing to electron-impurity scattering is given by
\begin{equation}
\nu_{im}= \frac{1}{\tau^{im}_{tr}} \int d \varepsilon \left(- \frac{\partial f(\varepsilon)}{\partial \varepsilon} \right) 
D^2({\varepsilon}) ,
\end{equation}
where $\tau^{im}_{tr}$ is the transport time for electron-impurity scattering. It is defined according 
to $1/\tau^{im}_{tr}=(m/2 \pi \hbar^3) \int_0^{2 \pi} d \theta~ w(2 p_F \sin\frac{\theta}{2}) (1-\cos \theta)$,
where $w$ is the Fourier transform of the random impurity-potential correlator. 
The self-consistent Born approximation leads to the density of states in the form 
of an expansion in oscillation harmonics weighted with powers of the Dingle factor 
$d \equiv \exp(-\pi/\omega_c \tau_q)$:$^{18,19}$
\begin{eqnarray}
D({\varepsilon})= 1 + 2 \sum_{k=1}^{\infty} a_k \cos\frac{2 \pi k \varepsilon}{\hbar \omega_c}, ~~~~~~~~~ \\ 
a_k=(-1)^k k^{-1} \exp(-\pi k/\omega_c \tau_q) L_{k-1}^1 (2 \pi k/\omega_c \tau_q ). \nonumber
\end{eqnarray} 
If the magnetic fields is weak, the Dingle factor is small. This corresponds to the case of 
overlapping Landau levels, when the density of states is given by a single-harmonic expression, 
$D(\varepsilon)=1-2d \cos(2 \pi \varepsilon/\hbar \omega_c)$. In strong enough magnetic fields the 
electron system is in the regime of separated Landau levels, when $D(\varepsilon)$ is represented 
by a periodic sequence of semi-elliptic peaks centered at the Landau quantization energies 
$\varepsilon_n$. 
 
The expressions (8)-(11), together with Eqs. (4)-(6) defining the phonon spectrum 
$\omega_{\lambda {\bf q}}$ and the function $C_{\lambda {\bf q}}$, give a complete description 
of the phonon-assisted 2D magnetotransport at large filling factors. The next step in 
evaluation of the resistivity can be done by calculating the integral over the energy 
$\varepsilon$. In this procedure, it is reasonable to omit the terms containing oscillating 
functions of energy under the integral. Such terms are responsible for the Shubnikov-de Haas 
oscillations, and they are exponentially suppressed with increasing temperature. Thus, the 
approximation used below corresponds to the condition
\begin{equation}
\frac{2 \pi^2 T/\hbar \omega_c}{\sinh(2 \pi^2 T/\hbar \omega_c)} \ll 1,
\end{equation}  
which means that the temperature should not be too low. As a result, 
\begin{eqnarray}
\nu_{ph}= \int_0^{2 \pi} \frac{d \varphi}{2 \pi} \int_0^{2 \pi} 
\frac{d \theta}{2 \pi} (1-\cos \theta) \nonumber \\
\times \int_{0}^{\infty} \frac{d u_z}{\pi} I\left(\frac{p_Fd_w}{\hbar} 
u_z \right) \sum_{\lambda} \nu_{\lambda {\bf q}} F\left(\frac{\hbar \omega_{\lambda {\bf q}}}{2T} \right) \nonumber \\  
\times \left[1+ 2  \sum_{k=1}^{\infty} a^2_k \cos \frac{2 \pi k \omega_{\lambda {\bf q}}}{\omega_c} \right] 
\equiv \nu_{ph}^{(0)} + 2  \sum_{k=1}^{\infty} a^2_k \nu_{ph}^{(k)}, 
\end{eqnarray}
where $F(x)=[x/\sinh(x)]^2$ and $u_z=\hbar q_z/2p_F$. The quantity 
\begin{equation}
\nu_{\lambda {\bf q}} = \frac{4 m T p_F C_{\lambda {\bf q}}}{\hbar^5 \omega_{\lambda {\bf q}} } 
\end{equation}
is an anisotropic scattering rate introduced for convenience purpose. The corresponding 
result for the impurity-assisted magnetotransport has a simpler form, $\nu_{im}=[1 + 
2 \sum_{k=1}^{\infty} a^2_k]/\tau^{im}_{tr}$, which describes a positive 
magnetoresistance oving to Landau quantization.$^{18}$ The Fermi momentum entering 
Eqs. (13) and (14) is related to electron density according to $p_F=\hbar \sqrt{2 \pi n_s}$.

\section{Results and discussion} 

The expression (13) is the central part of this paper. The first term of this expression,
$\nu_{ph}^{(0)}$, is responsible for phonon-induced contribution to background resistivity, 
while the second one corresponds to oscillations of the resistivity as a function of the 
magnetic field. These oscillations (PIRO) are described in Eq. (13) through a sum of oscillatory 
harmonics of the scattering rate, $\nu_{ph}^{(k)}$. Because of the complicated form of phonon spectrum, 
the integrals in Eq. (13) cannot be calculated analytically in the general case. However, an analytical
consideration is possible for the oscillating part of $\nu_{ph}$, since this part contains rapidly 
varying functions of $\omega_{\lambda {\bf q}}$ under the integral. To carry out the approximate 
integration, it is convenient to represent expressions for oscillatory harmonics by using the spherical 
coordinate system according to $\sin(\theta/2)=u \sin\chi$, $u_z=u \cos\chi$, where $u=\hbar q/2p_F$ is the 
absolute value of phonon wave vector in the dimensionless form and $\chi$ is the inclination angle. 
In these variables, the phonon spectrum is conveniently written as $\omega_{\lambda {\bf q}} = 
s^{(\lambda)}_{\varphi \chi} q$, where $s^{(\lambda)}_{\varphi \chi}$ is the anisotropic sound 
velocity for the mode $\lambda$. The angles $\varphi$ and $\chi$ determine the direction of 
the phonon wave vector. As a result,  
\begin{eqnarray}
\nu_{ph}^{(k)}= \frac{2}{\pi^2} \sum_{\lambda}
\int_0^{2 \pi} \frac{d \varphi}{2 \pi} \int_{0}^{\pi} d \chi \int_0^{(\sin\chi)^{-1}} d u \nonumber \\
\times \frac{u^3 \sin^2\chi}{\sqrt{1-u^2\sin^2\chi}} I\left( \frac{p_Fd_w}{\hbar} u \cos\chi \right) 
F\left(\frac{p_F s^{(\lambda)}_{\varphi \chi} u}{T} \right) \nonumber \\
\times \left[\nu^{DA(\lambda)}_{\varphi \chi} + u^{-2} \nu^{PA(\lambda)}_{\varphi \chi} \right]
\cos \frac{4 \pi k p_F s^{(\lambda)}_{\varphi \chi} u}{\hbar \omega_c},
\end{eqnarray}
where $\nu_{\lambda {\bf q}}$ given by Eq. (14) is written as the expression in the 
square brackets; the deformation-potential (DA) and piezoelectric (PA) contributions are separated according 
to Eq. (4). The main contribution to the integral over $u$ comes from the region $u \simeq (\sin\chi)^{-1}$, 
and the convergence of this integral is determined by the rapidly oscillating cosine function. The 
other $u$-dependent functions, $I$ and $F$, are slowly varying on the scale of convergence because 
of $p_F d_w/\hbar \sim 1$ and because of condition (12), respectively. After integrating over $u$, 
one obtains
\begin{eqnarray}
\nu_{ph}^{(k)}= \frac{1}{\pi^2} \sum_{\lambda}
\int_0^{2 \pi} \frac{d \varphi}{2 \pi} \int_{0}^{\pi} d \chi  \nonumber \\
\times \sqrt{\frac{\omega_c \sin \chi}{2 k p_F s^{(\lambda)}_{\varphi \chi}}}I\left(\frac{p_Fd_w}{\hbar} 
\cot\chi \right) F\left(\frac{p_F s^{(\lambda)}_{\varphi \chi}}{T\sin \chi} \right) \nonumber \\
\times \left[\nu^{DA(\lambda)}_{\varphi \chi}/\sin^2\chi + \nu^{PA(\lambda)}_{\varphi \chi} \right]
\cos \left[ \frac{4 \pi k p_F s^{(\lambda)}_{\varphi \chi}}{\hbar \omega_c \sin \chi} -\frac{\pi}{4} \right].
\end{eqnarray}
The remaining double integral over angular variables can be calculated using the method of 
fastest descent. Indeed, under the integral we have a rapidly oscillating function (the argument 
of the cosine in Eq. (16) is assumed to be large), and the main contribution 
comes from certain regions of $\varphi$ and $\chi$ where $\Phi^{(\lambda)}_{\varphi \chi} 
\equiv s^{(\lambda)}_{\varphi \chi}/ \sin \chi$ varies most slowly. Naturally, these regions are 
in the close vicinity of the extrema points of the function $\Phi^{(\lambda)}_{\varphi \chi}$. The 
anisotropic sound velocity $s^{(\lambda)}_{\varphi \chi}$ itself has a number of local maxima, minima, 
and saddle points, whose positions coincide with some high-symmetry directions in the reciprocal 
(momentum) space. The number of extrema for $\Phi^{(\lambda)}_{\varphi \chi}$ is smaller because 
of the factor $1/\sin \chi$. For the mode with the highest velocity (this mode becomes purely 
longitudinal for high-symmetry directions) the anisotropy is not strong, and all the extrema 
are only at $\chi=\pi/2$ (zero $q_z$). The slow modes (which become purely transverse for 
high-symmetry directions) also have extrema at $\chi=\pi/2$. Since the anisotropy of these 
modes is stronger, they also may have additional groups of extrema at $\chi=\pi/2 \pm \xi$, 
where $\xi <\pi/4$. Nevertheless, the main contribution of slow (low-energy) modes to oscillating
magnetoresistance is associated with the extremum $\chi=\pi/2$ for a single mode whose velocity 
at $\chi=\pi/2$ is independent of the polar angle $\varphi$. This mode at $\chi=\pi/2$ is purely 
transverse and polarized perpendicular to the quantum well plane (direction [001]). 

In summary, the analysis shows that the modes effectively contributing to the oscillating part of 
resistivity are: i) the transverse mode polarized along [001], whose velocity $s_{T0}=\sqrt{c_{44}/\rho}$ 
is independent of $\varphi$ and corresponds to a maximum of $\Phi^{(\lambda)}_{\varphi \chi}$ 
as a function of $\chi$; ii) the longitudinal mode with velocity $s_{L0}= \sqrt{c_{11}/\rho}$ 
polarized along [100] or along equivalent directions ([010], [$\bar{1}$00], and [0$\bar{1}$0]) 
corresponding to local minima of $\Phi^{(\lambda)}_{\varphi \chi}$; 
iii) the longitudinal mode with velocity $s_{L1}=\sqrt{(c_{11}+c_{12}+2c_{44})/2\rho}$ 
polarized along [110] or along equivalent directions ([1$\bar{1}$0], [$\bar{1}$10], 
and [$\bar{1} \bar{1}$0]) corresponding to saddle points of $\Phi^{(\lambda)}_{\varphi \chi}$. 
Accordingly, the oscillatory harmonics of the scattering rate are written as sums of three 
components:
\begin{equation}
\nu_{ph}^{(k)}=\nu_{T0}^{(k)}+ \nu_{L0}^{(k)} + \nu_{L1}^{(k)}, 
\end{equation}
where 
\begin{eqnarray}
\nu_{T0}^{(k)} = \frac{(eh_{14})^2 mT \omega_c}{8\pi^2 \hbar k b_{T0} \rho s_{T0}^3 p_F^2} 
F\left(\frac{p_F s_{T0}}{T} \right) \sin \frac{4 \pi k p_F s_{T0}}{\hbar \omega_c},
\end{eqnarray}
\begin{eqnarray}
\nu_{L0}^{(k)} = \frac{\sqrt{2} {\cal D}^2 mT \omega_c^{3/2}}{\pi^3 \hbar^{5/2} k^{3/2} b_{L0} \rho s_{L0}^{7/2} 
p_F^{1/2}} \nonumber \\ \times F\left(\frac{p_F s_{L0}}{T} \right) \cos \left( \frac{4 \pi k p_F s_{L0}}{\hbar \omega_c} + \frac{\pi}{4} \right),
\end{eqnarray}
\begin{eqnarray}
\nu_{L1}^{(k)} = \frac{\sqrt{2} {\cal D}^2 mT \omega_c^{3/2}}{\pi^3 \hbar^{5/2} k^{3/2} b_{L1} \rho s_{L1}^{7/2} 
p_F^{1/2}} \nonumber \\ \times F\left(\frac{p_F s_{L1}}{T} \right) \cos \left( \frac{4 \pi k p_F s_{L1}}{\hbar \omega_c} - \frac{\pi}{4} \right).
\end{eqnarray}
The coefficients of order unity standing in the denominators of these expressions are
given by
\begin{equation}
b_{T0}^2=-1 + \frac{(c_{11}+c_{44}+2c_{12})c_A}{c_{44}(c_{11}+c_{12})},
\end{equation}
\begin{eqnarray}
b_{L0}^2=(1 + \gamma)\gamma,~~\gamma=\frac{(c_{11}+c_{12}) c_A}{c_{11}(c_{11}-c_{44})},
\end{eqnarray}
and
\begin{eqnarray}
b_{L1}^2=(1 + \beta_1)\beta_2,~~~~~\\ 
\beta_1=\frac{(3c_{12}+2c_{44}+c_{11})c_A}{(c_{11}+c_{12}+2c_{44})(c_{11}+c_{12})}, \nonumber \\
\beta_2=\frac{(c_{11}+c_{12}) c_A}{(c_{11}+c_{12}+2c_{44})(c_{44}+c_{12})}, \nonumber 
\end{eqnarray}
where $c_A=2c_{44}+c_{12}-c_{11}$ is the positive quantity characterizing the 
anisotropy of phonon spectrum. The contribution from transverse phonons is associated with 
piezoelectric potential, while the contribution from longitudinal phonons is due to 
deformation potential of electron-phonon interaction. 

The accuracy of the analytical approach leading to Eqs. (17)-(20) is not expected to be high 
for lowest-order PIRO peaks. Nevertheless, the applicability of these equations is steadily improved 
as one moves to lower magnetic fields and the argument of the oscillating function becomes 
larger. For such low fields, the rate $\nu_{T0}^{(k)}$ tends to overcome the rates 
$\nu_{L0}^{(k)}$ and $\nu_{L1}^{(k)}$. However, in the samples with higher electron density, 
where the characteristic phonon wave number $2p_F/\hbar$ is larger, the deformation-potential 
mechanism is more significant than the piezoelectric one. Therefore, both transverse-phonon and 
longitudinal-phonon contributions are important and should be taken into account. Notice also 
that for low magnetic fields corresponding to the case of overlapping Landau levels one can 
take only the lowest harmonic $k=1$ in the sum in Eq. (13). This reduces the oscillating part 
of phonon-induced transport rate $\nu_{ph}$ to a simpler form $2 d^2 \nu_{ph}^{(1)}$.

For completeness, one can also present the result obtained in the isotropic approximation, 
where there are a longitudinal mode and two degenerate transverse modes with velocities $s_{l}$ 
and $s_{t}$, respectively:
\begin{eqnarray}
\nu_{ph}^{(k)}=\frac{(eh_{14})^2 mT \omega_c}{8\pi^2 \hbar k \rho s_{t}^3 p_F^2} 
F\left(\frac{p_F s_{t}}{T} \right) \cos \frac{4 \pi k p_F s_{t}}{\hbar \omega_c} \nonumber \\
+ \frac{{\cal D}^2 m T \omega_c}{\pi^2 \hbar^3 k \rho s_{l}^3}
F\left(\frac{p_F s_{l}}{T} \right) \cos \frac{4 \pi k p_F s_{l}}{\hbar \omega_c}. 
\end{eqnarray}
In this approximation, the phase of the oscillations, as well as the dependence on magnetic 
field and Fermi momentum is different from the case of anisotropic phonon spectrum.   

\begin{figure}[ht]
\includegraphics[width=9.2cm]{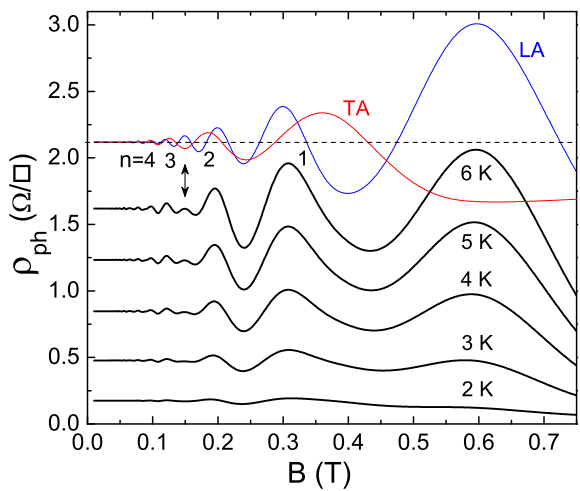}
\caption{(Color online) Phonon-induced resistivity calculated for the quantum 
well of experiment Ref. 17 at temperatures from 2 to 6 K. For $T=6$ K, the partial 
oscillating contributions from high-energy branch (LA) and low-energy branches (TA) 
are also shown (shifted for clarity).}
\end{figure}
  
The numerical calculations described below use the following elastic constants for GaAs 
(in units $10^{11}$ dyn/cm$^2$): $c_{11}=12.17-0.00144~T$(K), $c_{12}=5.46-0.00064~T$(K), and
$c_{44}=6.16-0.0007~T$(K), taken from a semiconductor material database.$^{20}$ The temperature 
dependence of these constants is not really essential in the range of $T$ corresponding to 
experimental studies of oscillating magnetoresistance. The other GaAs parameters used are 
${\cal D}=7.17$ eV, $h_{14}=1.2$ V/nm, and $\rho=5.317$ g/cm$^3$. An example of the results 
of numerical calculations according to Eq. (13) is presented in Fig. 1. The calculation 
employs the following parameters of the high-mobility quantum well studied in Ref. 17: 
$n_s=3.75 \times 10^{11}$ cm$^{-2}$, $d_w=30$ nm, impurity-limited mobility $1.17 
\times 10^{7}$ cm$^2$/V s, and quantum lifetime owing to impurity scattering $\tau_q=15$ ps. 
The theoretical plots are in agreement with experimental data of Ref. 17 as concerns 
the background resistance, PIRO peak positions, and the amplitude of the oscillations 
(for example, at $T=4$ K the amplitudes of the peaks numbered 1 and 2 are 0.16 and 0.075 Ohm 
per square, respectively). The calculation shows that the phonon-induced resistivity is determined 
by both longitudinal and transverse modes, whose contributions are nearly equal to each other at 
the fields $B > 0.1$ T and at 6 K. At lower fields the transverse-mode contribution becomes more 
important. With decreasing temperature both contributions are suppressed according to the  
factor $T F(\hbar \omega_{\lambda {\bf q}}/2T)$. The LA-mode contribution is suppressed 
stronger because the thermal activation energy for this mode is larger. Figure 1 shows that 
the peak placed slightly below 0.6 T is the lowest-order one for the LA-mode contribution, 
so the origin of this peak (discussed in Ref. 17) seems to be clear. However, the relative 
amplitude of this peak is higher than that observed in the experiment, which is possibly related 
to overestimation of the deformation-potential interaction in comparison to the piezoelectric 
one.$^{21}$ The other low-order peaks ($n=1$, 2, 3) are formed by both transverse-mode and 
longitudinal-mode contributions, so their positions $B_n$ are not expected to follow exactly a 
$1/B$-periodic dependence. Nevertheless, the deviations appear to be small, and a 9-point linear fit 
of theoretical peak positions (see Fig. 2) gives the dependence $n=2 p_F s/\hbar \omega_c - 
\delta n$, where $s \simeq 3.44$ km/s is very close both to the velocity $s_{T0} \simeq 3.40$ 
km/s and to the velocity found experimentally,$^{17}$ while $\delta n \simeq 0.21$ is 
slightly larger than that obtained from the experiment.$^{17}$ The parameter $\delta n$ 
determines the phase of the magnetoresistance oscillations, which appears to be close to the 
phase of the oscillating factor in the approximate analytical expression (18), $\delta n=1/4$. 
Starting from $T=3$ K, the theoretical dependence has a weak extra peak (marked by the arrow in Fig. 1) 
at $B \simeq 0.15$ T, which comes from the longitudinal-mode contribution. Though the 
experimental magnetoresistance$^{17}$ does not show a peak in this region, a precursor 
of such a feature is apparently visible as a flattening of the minimum between the peaks 
$n=2$ and $n=3$ at the temperatures above 4 K. 

\begin{figure}[ht]
\includegraphics[width=8cm]{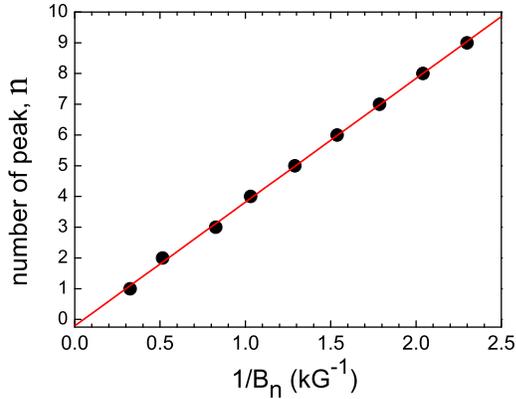}
\caption{(Color online) Peak number $n$ as a function of inverse peak 
position $1/B_n$ extracted from the calculated magnetoresistance (Fig. 1) at 4 K. 
A linear fit for this dependence yields $s \simeq 3.44$ km/s. The same kind of plot 
based on the experimental data is given in Ref. 17. The positions of low-order 
peaks slightly deviate from the linear dependence because these peaks are 
formed by mixture of the oscillating contributions from different phonon modes.}
\end{figure}

It is important to compare the results of numerical calculation of the oscillating 
component of phonon-induced magnetoresistance with the results obtained from the 
analytical approach leading to Eqs. (17)-(20). Figure 3 (a) shows the partial contribution 
of high-energy phonon mode (LA) calculated according to Eq. (13) together with the corresponding 
contribution based on the expression $\nu_{ph}^{(k)}=\nu_{L0}^{(k)}+\nu_{L1}^{(k)}$, 
where $\nu_{L0}^{(k)}$ and $\nu_{L1}^{(k)}$ are given by Eqs. (19) and (20). In a similar way, 
Fig. 3 (b) shows both numerical and analytical, based on Eq. (18), results for low-energy 
phonon modes (TA). The agreement between exact and approximate results becomes good in the 
low-field region corresponding to high-order magnetophonon resonances. Both numerical and 
approximate results in Fig. 3 (a) demonstrate a beating pattern with a node at $B \simeq 0.08$ T. 
The analytical consideration reveals the origin of this beating, since it shows that the 
oscillations induced by LA phonons are formed as a sum of two contributions, $\nu_{L0}^{(k)}$ 
and $\nu_{L1}^{(k)}$, with slightly different frequencies.

\begin{figure}[ht]
\includegraphics[width=9.2cm]{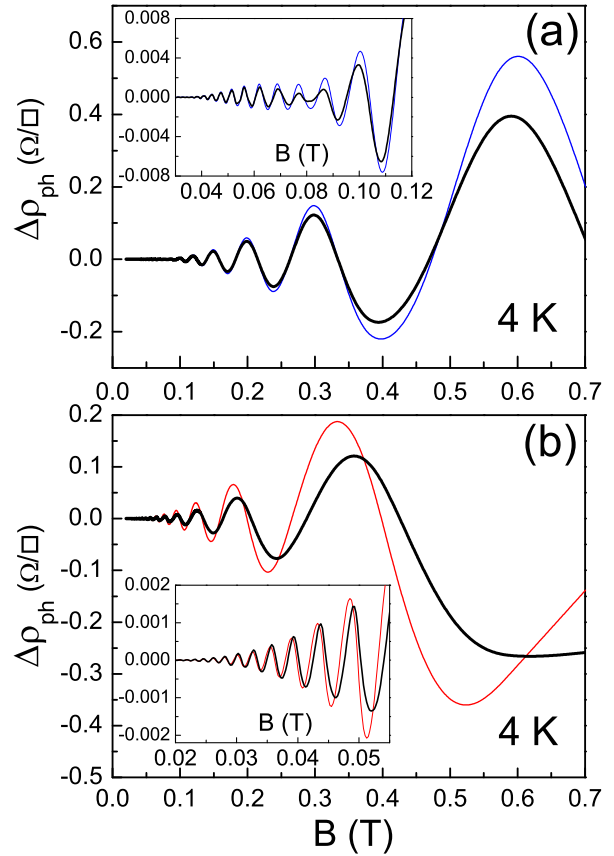}
\caption{(Color online) Partial oscillating contributions of the phonon-induced 
resistivity for the quantum well of experiment Ref. 17 at $T=4$ K. (a) Contribution of 
high-energy branch (LA). Thick (black) lines correspond to numerical calculation of $\nu_{ph}^{(k)}$ 
using Eq. (13), and thin (blue) lines are the results obtained from the approximate analytical 
expressions (19) and (20). The inset shows beating pattern in the low-field region. 
(b) Contribution of low-energy branches (TA). Thick (black) and thin (red) lines correspond to 
numerical calculation using Eq. (13) and to approximate analytical expression (18), 
respectively. The low-field region is shown in the inset.}
\end{figure}

Figure 4 presents the phonon-induced magnetoresistance calculated for the GaAs quantum wells 
with $n_s=10^{12}$ cm$^{-2}$ and $d_w=13$ nm studied in Ref. 14. Since the mobility of 
these samples ($\simeq 6.6 \times 10^5$ cm$^2$/V s at $T=4.2$ K according to the resistance 
at $B=0$) is much lower than that in Ref. 17, both experimental$^{14}$ and theoretical 
magnetoresistance show a few low-order oscillations. Because of large density and, 
consequently, large Fermi momentum in these samples, the main contribution to oscillating 
resistivity is caused by the deformation-potential interaction and comes from LA 
phonons. This is demonstrated by plotting partial contributions of different modes 
in Fig. 4. At the temperatures 15-25 K all acoustic-phonon modes are fully activated, 
which means that the function $F(\hbar \omega_{\lambda {\bf q}}/2T)$ is close to 1, so 
the resistivity should linearly increase with temperature. Experimental plots of Ref. 14 
indeed show a linear increase for the background resistivity, while the amplitudes of 
the oscillation peaks depend on $T$ in a different way: the first peak is weakly modified 
by $T$ and the second peak is suppressed with increasing $T$. Such a behavior can be 
explained by a decrease of the quantum lifetime with increasing temperature owing to 
inelastic electron-electron scattering (see Ref. 17 and references therein). In the 
present formalism, this effect is described by adding the electron-electron scattering 
rate $1/\tau_q^{ee}= \lambda T^2/\hbar \varepsilon_F$, where $\lambda \sim 1$, to the 
inverse quantum lifetime$^{22}$ $1/\tau_q$ entering the density of states in Eq. (11). 
The result of the calculations improved in this way is also shown in Fig. 4. A reasonably 
good agreement with experimental data, as concerns the amplitudes of the oscillations and 
their temperature dependence, is achieved at $\lambda=2$. 

\begin{figure}[ht]
\includegraphics[width=9.2cm]{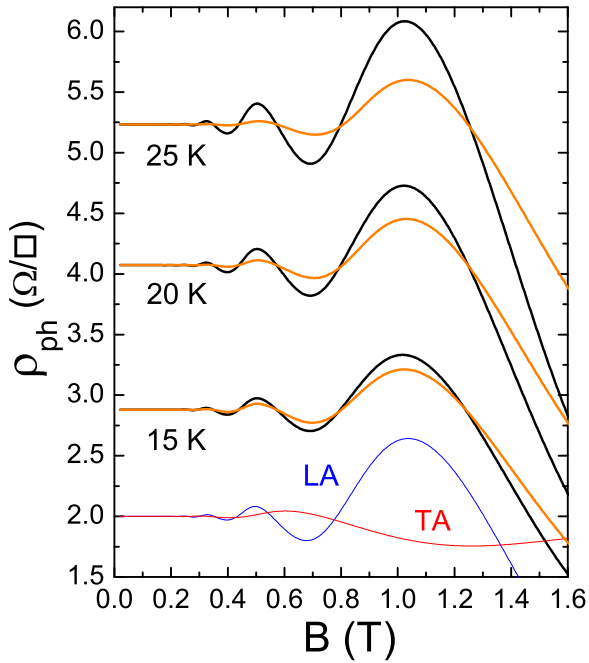}
\caption{(Color online) Phonon-induced resistivity calculated for 
the quantum well of experiment Ref. 14 at different temperatures. For $T=15$ K, the 
partial oscillating contributions from high-energy branch (LA) and low-energy branches (TA) 
are also shown (shifted for clarity). The additional (orange) plots take into 
account temperature-induced decrease of the quantum lifetime and are in a better 
agreement with experimental data of Ref. 14.}
\end{figure}

The positions of the calculated magnetoresistance peaks deviate from the experimental 
data, though this deviation is not strong. The fitting procedure similar to that shown 
in Fig. 2 gives the characteristic velocity $s=5.1$ km/s, which is smaller than $s=5.9$ km/s 
determined in Ref. 14. It is worth noting, however, that the positions of the peaks in the 
experiment Ref. 14 are shifted to higher magnetic fields as the temperature increases. 
Though the reason of this shift is not clear, it signifies an increase of experimental 
velocity $s$ with temperature. Therefore, one may expect a better agreement of theory 
and experiment at lower temperatures. 

\section{Summary and conclusions}

This paper presents a microscopic theory of magnetophonon oscillations, also 
known as phonon-induced resistance oscillations (PIRO), in 2D electron systems. 
The phonon-induced resistivity $\rho_{ph}$ is found by considering both 
deformation-potential and piezoelectric mechanisms of interaction of 2D electrons 
with bulk acoustic phonon modes. In the experimentally relevant situation of weak 
magnetic fields (large filling factors), the expression for the resistivity (or, 
eqivalently, for the phonon-assisted transport rate $\nu_{ph}$) is determined by 
the electron-phonon coupling constants, phonon frequencies, and the density of 
electron states in magnetic field; see Eq. (13). The essential feature of the 
calculations is that the anisotropy of the phonon spectrum is explicitly 
taken into account. The theory is applied to quantum wells based on the materials of 
cubic symmetry and grown in the [001] crystallographic direction. For this case, an
analytical expression for the oscillating part of the resistivity has been derived. 
This expression is valid in the region of fields corresponding to high-order 
magnetophonon resonances, when the ratio of characteristic phonon frequency to 
cyclotron frequency is large. The calculations are carried out for GaAs quantum 
wells where PIRO have been observed experimentally. This includes the cases of 2D 
electron gas with high density and moderate mobility$^{14}$ and with lower density 
and very high mobility.$^{17}$ The calculated magnetoresistance is in agreement 
with experimental data. As concerns the positions of the magnetoresistance peaks, 
the agreement is very good in the case of high-mobility systems.$^{17}$  

Based on the results obtained, one may conclude that the bulk approximation for 
description of acoustic phonon modes and of their interaction with 2D electrons works
reasonably good in application to magnetotransport in GaAs quantum wells. Another 
important conclusion is that the velocity $s$ in the empirical equation (1) does 
not, in general, correspond to a certain phonon mode. The reason for this is the
complicated structure of the phonon spectrum. There are three anisotropic phonon
modes (branches) with different velocities and two mechanisms, deformation-potential 
and piezoelectric, of electron-phonon interaction. Relative contributions of these 
modes and of the interaction mechanisms depend on 2D electron density, temperature, and 
other parameters such as the width of the quantum well and growth direction. 
This explains why the velocity $s$ determined by fitting the experimental PIRO peak 
positions is expected to vary in different experiments, especially when such 
a fit is based upon a few low-order peaks which have the highest amplitudes 
and, therefore, are best visible experimentally. Indeed, the low-order peaks 
are typically formed as a superposition of contributions from different modes, 
and even for each single mode there is a mixture of phonons with different 
velocities owing to the anisotropy. The anisotropy influences not only the 
frequency but also the phase and the amplitude of the oscillations.

On the other hand, in the region of high-order magnetophonon resonances 
the resistance oscillations are described by a fixed set of phonon 
velocities determined here for the case of [001]-grown quantum wells; 
see Eqs. (17)-(20). The phases of these oscillations also become definite. 
Moreover, if the density of 2D electrons is not large, so the piezoelectric-potential 
interaction is significant, the main contribution to magnetoresistance oscillations 
in [001]-grown wells is characterized by a single velocity $s=s_{T0} \equiv 
\sqrt{c_{44}/\rho}$ corresponding to a TA mode propagating in the quantum 
well plane and polarized perpendicular to this plane. This is the case of 
experiment Ref. 17, where many magnetoresistance oscillations have been 
observed owing to a very high electron mobility. For the wells whose growth axes are 
different from [001] the characteristic phonon velocities have to be different. 
These cases are not studied here in detail because PIRO have not yet been 
observed in such wells. Nevertheless, based on the consideration presented in 
Sec. III, the following general method for determination of characteristic 
phonon velocities in the wells of arbitrary growth axis is proposed. Take the 
surfaces of equal frequency $\omega$ defined for each mode in the ${\bf q}$-space 
by the equation $s^{(\lambda)}_{\varphi \chi} q =\omega$. Find the points 
(numbered by the index $i$) where these surfaces touch the surface of the cylinder of 
radius $2p_F/\hbar$ and axis along the growth direction (this cylindrical surface is 
given by the equation $q \sin \chi=2p_F/\hbar$). The set of frequencies $\omega=
\omega_{\lambda i}$ (or, equivalently, velocities $s_{\lambda i}=\omega_{\lambda i}/q$) 
corresponding to these points will give a set of oscillating harmonics with 
arguments $2 \pi k \omega_{\lambda i}/\omega_c$ in the magnetoresistance. An 
equivalent procedure is the search for extrema of the function $s^{(\lambda)}_{\varphi 
\chi}/\sin \chi$, as described in Sec. III. To find which frequencies out of 
$\omega_{\lambda i}$ give the main contribution and to determine the phases 
and amplitudes of the oscillations, a detailed analysis is necessary. 
In the case of [001]-grown wells a single TA mode gives the main contribution [Eq. (18)] 
because the surface of equal frequency for this particular mode touches 
the cylinder not just in a finite number of points, but along the whole circumference 
of the cylinder at $\chi=\pi/2$. 

To study the influence of different phonon modes on the magnetoresistance oscillations, 
it is desirable to carry out experiments in high-mobility 2D systems with densities
varied in the range $4-10 \times 10^{11}$ cm$^{-2}$. If it is possible to reach high 
mobilities for the structures with growth axes different from [001], measurements 
of the oscillating magnetoresistance in such systems would be also important for 
investigation of the phonon anisotropy effects discussed above. The author hopes 
that the present theoretical work may stimulate such experiments. 

The consideration given in this paper assumes the linear response regime, when 
both electron and phonon systems are close to equilibrium. One may expect a more 
interesting and rich behavior of the phonon-induced resistance oscillations 
under nonlinear transport regime when the electric current through 
the sample increases. Studies in this direction are already undertaken.$^{15,16}$ In addition 
to the work already done, more efforts, both experimental and theoretical, are necessary for 
better understanding of the mechanisms of phonon-induced oscillations and their
interplay with the other microscopic mechanisms responsible for the magnetoresistance 
of 2D electrons at large filling factors.


\begin{thebibliography}{22}
\bibitem{22}
Phonon-assisted scattering also leads to tempearature 
dependence of quantum lifetime. However, the corresponding additional 
contribution to the inverse quantum lifetime is of the order of 
background phonon-assisted transport rate $\nu_{ph}^{(0)}$ and 
is much smaller than $1/\tau_q$ caused by electron-impurity 
scattering.
\end{thebibliography}
\end{document}